\documentclass[manuscript]{emulateapj}

\usepackage{multirow}
\usepackage{graphicx}
\usepackage{rotating}
\usepackage{amsmath, amsthm, amssymb}
\usepackage{color}

\begin{document}

\title{Magnetic Fields and Galactic Star Formation Rates}
\author{Sven Van Loo}
\affil{School of Physics and Astronomy, University of Leeds, Leeds LS2 9JT, UK}
\affil{Harvard-Smithsonian Center for Astrophysics, 60 Garden Street, Cambridge, MA 02138, USA}
\email{S.VanLoo@leeds.ac.uk}
\author{Jonathan C. Tan}
\affil{Departments of Astronomy and Physics, University of Florida, Gainesville, FL 32611, USA}
\author{Sam A. E. G. Falle}
\affil{Department of Applied Mathematics, University of Leeds, Leeds LS2 9JT, UK}

\begin{abstract}
The regulation of galactic-scale star formation rates (SFRs) is a
basic problem for theories of galaxy formation and evolution: which
processes are responsible for making observed star formation rates so
inefficient compared to maximal rates of gas content divided by
dynamical timescale? Here we study the effect of magnetic fields of
different strengths on the evolution of giant molecular clouds (GMCs)
within a kiloparsec patch of a disk galaxy and resolving scales down
to $\simeq0.5\:{\rm{pc}}$. Including an empirically motivated
prescription for star formation from dense gas
($n_{\rm{H}}>10^5\:{\rm{cm}^{-3}}$) at an efficiency of 2\% per local
free-fall time, we derive the amount of suppression of star formation
by magnetic fields compared to the nonmagnetized case. We find GMC
fragmentation, dense clump formation and SFR can be significantly
affected by the inclusion of magnetic fields, especially in our
strongest investigated $B$-field case of $80\:{\rm{\mu}}$G. However,
our chosen kpc-scale region, extracted from a global galaxy
simulation, happens to contain a starbursting cloud complex that is
only modestly affected by these magnetic fields and likely requires
internal star formation feedback to regulate its SFR.
\end{abstract}


\maketitle

\section{Introduction}

Understanding and thus predicting the star formation rate (SFR) that
results from a galactic disk of given local properties, such as gas
mass surface density, $\Sigma_g$, and orbital timescale, is a
necessary foundation on which to build theories of galaxy
evolution. Global and kiloparsec-scale correlations between star
formation activity, gas content, and galactic dynamical properties
have been observed \citep[][]{KennicuttEvans2012}. In molecular-rich
regions of normal disk galaxies, overall star formation rates are
relatively slow and inefficient, i.e., a small fraction,
$\epsilon_{\rm{orb}}\simeq0.04$, of total gas is converted to stars
every local galactic orbital time \citep[][]{Suwannajak_ea2014}. Most
star formation is concentrated within $\sim1-10\:{\rm{pc}}$-scale
regions within GMCs, but even here star formation appears to be
inefficient, with similarly small fractions,
$\epsilon_{\rm{ff}}\simeq0.01-0.05$, of total gas forming stars every
local free-fall time, $t_{\rm{ff}}\equiv(3\pi/[32{G}\rho])^{1/2}$,
where $\rho$ is gas density
\citep[][]{KrumholzTan2007,DaRio_ea2014}. Low efficiencies of star
formation may be the result of some combination of turbulence
\citep[][]{Padoan_ea2014}, magnetic fields
\citep[][]{McKee1989,TassisMouschovias2004} and star formation
feedback \citep[][]{Krumholz_ea2014}. Here we consider the effects of
magnetic fields on the evolution, collapse and SFR of a turbulent,
shearing, kiloparsec-scale region of a galactic disk, extracted from
the global galaxy 
simulation of \citet[][hereafter
  TT09]{TaskerTan2009}.  

\citet[][hereafter Paper I]{VLBT2013} showed that in the limit of
purely hydrodynamic evolution with no star formation feedback, GMCs
forming, evolving and collapsing in this same kiloparsec-scale
environment produce too much high density gas and, hence, the SFR is
too high. Additional processes are needed to provide extra support to
or disruption of the GMCs.  One such process may be support by
magnetic fields.  Observations show that clouds such as, e.g., Taurus
and $\rho$Oph, are threaded by strong $B$-fields,
\citep[e.g.,][]{Heiles2000,Goldsmith_ea2008} and that the field is
connected on larger scales to galactic fields
\citep[][]{LiHenning2011}. The average solar-neighborhood Galactic
magnetic field is $6\pm2\:{\rm{\mu}{G}}$ \citep[][]{Beck2001}.
\citet[][]{Crutcher_ea2010} argue for a random distribution of field
strengths from 0 to $B_{\rm{max}}$ with
$B_{\rm{max}}=B_0=10\:{\rm{\mu}{G}}$ in low density gas
($n_{\rm{H}}<300\:{\rm{cm}^{-3}}$) and
$B_{\rm{max}}=B_0(n_{\rm{H}}/300\:{\rm{cm}^{-3}})^{0.65}$ at higher
denisties.  Dynamo amplification to equipartition values is a natural
expectation, which has also been seen in numerical simulations
\citep[e.g.,][]{WangAbel2009,PakmorSpringel2013}. The magnitude and
direction of the $B$-field may play an important role in the formation
of molecular clouds that are forming from compression of more diffuse
atomic gas \citep[][]{Heitsch_ea2009,VanLoo_ea2010}. By potentially
stabilizing and supporting GMCs, they may increase cloud lifetimes to
the point that other processes such as GMC collisions
\citep[][]{Tan2000} or spiral arm passage \citep[][]{Bonnell_ea2013}
become important.

Here, we extend the model presented in Paper~I to include $B$-fields,
varying their strength to investigate the effect on GMC evolution,
especially the amount of high-density clump gas and SFR.
\S\ref{sect:model} describes the numerical model, \S\ref{sect:results}
presents the results, and \S\ref{sect:summary} summarizes and
discusses their implications.

\section{Numerical model}\label{sect:model}

We start with the same initial conditions 
as Paper~I, i.e., a kiloparsec-sized box at $4.25\:{\rm{kpc}}$ from
the galactic center extracted from the TT09 simulation. We include the
same physical processes, i.e., heating/cooling functions developed in
Paper~I, and adopt a static background potential yielding flat
rotation curve of $200\:{\rm{km\:s^{-1}}}$.  As the TT09 simulation is 
nonmagnetic, no self-consistent initial magnetic field configuration is 
available. Therefore, for simplicity, we
thread the domain with uniform $B$-field along the shearing direction,
i.e., $\bf{B}=B_{0}\hat{1}_y$.  For $B_0$ we assume 0 (no $B$-field),
10 or 80$\:{\rm{\mu}}$G. The 10$\:{\rm{\mu}}$G case represents field
strengths close to mean Galactic values expected at an inner
($\sim4$~kpc) location. The 80$\:{\rm{\mu}}$G case 
exceeds observed kpc-scale $B$-fields by $\sim1\:$dex, but allows for
stronger fields inside GMCs and their substructures, which are already
present in the initial conditions.  Using the empirical relation of
\citet[][]{Crutcher_ea2010}, a field strength of
$B_{\rm{max}}=80\:{\rm{\mu}}$G corresponds to a density of
$n_{\rm{H}}\simeq7000\:{\rm{cm}^{-3}}$. The mean density of the two
most massive GMCs identified in Paper I is $\simeq470$ and
$330\:{\rm{cm}}^{-3}$, but they contain substructures extending to
more than ten times higher densities. Thus our adopted high magnetic
field case may be more appropriate for predicting the dynamical
evolution of these dense clumps and filaments.  The above field
strengths can also be compared to the critical values needed to
support idealized self-gravitating clouds of a given mass surface
density
$B_{\rm{crit}}=21.6\:\Sigma/(0.01\:{\rm{g\:cm^{-2}}})\:{\rm{\mu}{G}}=45.1\:\Sigma/(100\:M_\odot{\rm{pc^{-2}}})\:{\rm{\mu}{G}}$
\citep[e.g.,][]{MouschoviasSpitzer1976,McKee1999}. The TT09 simulation GMCs have typical
$\Sigma\sim{\rm few}\times100\:M_\odot{\rm{pc}^{-2}}$, so only in the
strongest field case do we expect significant influence on global GMC
dynamics.  Indeed, only in this 80$\:{\rm{\mu}}$G case are initial 
mass-to-flux ratios of the three least massive GMCs below the critical value.
Otherwise, the GMCs are supercritical by at least an order of
magnitude.  A larger number of intermediate $B$-field strengths and
different initial geometries will be explored in a future paper.

\begin{figure*}
\begin{center}
\includegraphics[width=\textwidth]{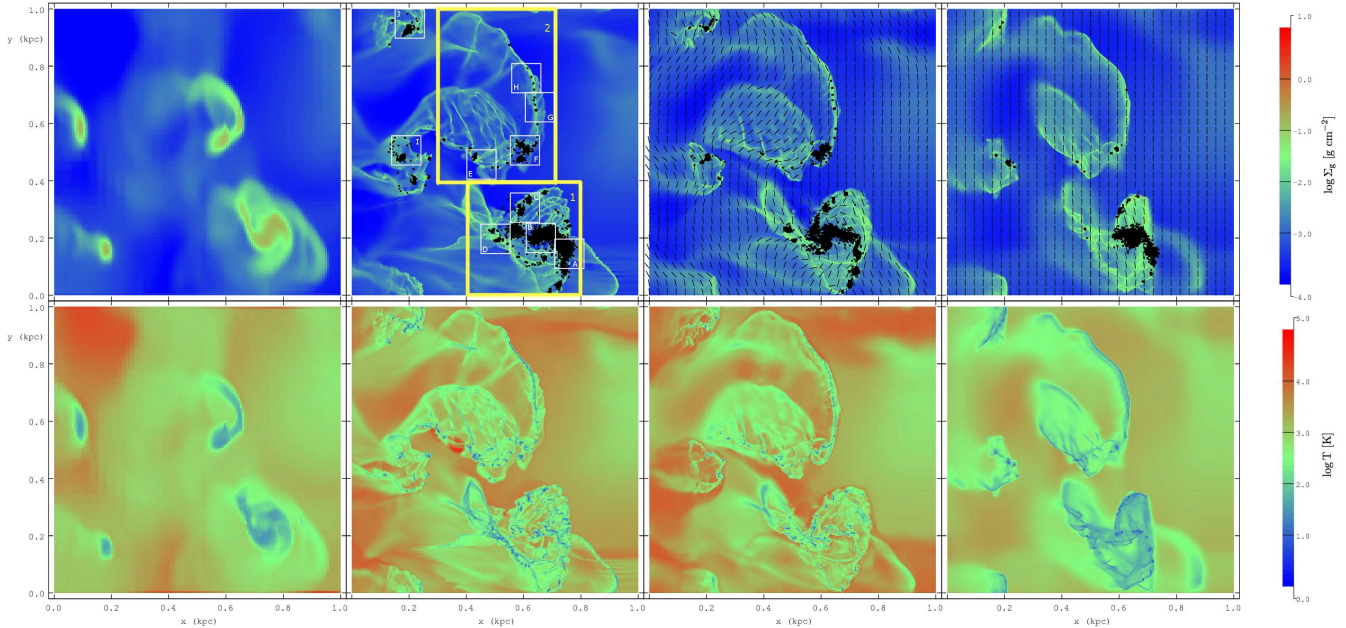}
\caption{
Logarithmic mass surface density 
(top) and logarithmic mass-weighted temperature 
(bottom) integrated along $z$-axis for (from left to right): initial
conditions ($t=0$); final conditions ($t=10\:$Myr) with
$B=0\:{\rm{\mu}}$G, $10\:{\rm{\mu}}$G and $80\:{\rm{\mu}}$G.
Lines indicate projected
direction of mass-weighted magnetic field and black dots show
positions of star cluster particles. Boxes in the second top panel
show different ``region'' (yellow) and ``cloud'' (white) selections.
}
\label{fig:surface_density}
\end{center}
\end{figure*}

To solve the ideal magnetohydrodynamics (MHD) equations we use the
Adaptive Mesh Refinement MHD code {\it{MG}}
\citep[][]{VanLoo_ea2006,Falle_ea2012}, rather than {\it{Enzo}}, which
was used in Paper I.
The basic algorithm is a second-order Godunov scheme with Local
Lax-Friedrichs solver and piecewise-linear reconstruction
method. Self-gravity is computed using a full approximation multigrid
to solve the Poisson equation. We also solve the non-conservative
internal energy equation to ensure positive pressures. We use the gas
temperature value from internal energy when this is $<1/10$ of total
energy and from total energy otherwise. To ensure that the solenoidal
constraint is met, divergence cleaning is implemented 
following \citet[][]{Dedner_ea2002}.

Refinement is on a cell-by-cell basis, controlled by the
\citet[][]{Truelove_ea1997} criterion. Note that, with the magnetic
Jeans length $\lambda_B=\lambda_J\sqrt{1+v_{A}^2/c_{s}^2}$ \citep[with
  $v_{A}$ the Alfv\'en speed and $c_s$ the sound speed;
  e.g.,][]{Strittmatter1966}, artificial fragmentation is less likely
when a magnetic field is included.  
We adopt the same boundary conditions (pseudo-shearing box) and
effective grid resolution of 2048$^2\times$4096 (by using a root grid
resolution of 128$^2\times$256 with 4 refinement levels) as
Paper~I. Again, we follow evolution for 10$\:\rm{Myr}$, less than one
shear-flow crossing time over the domain.

Inclusion of a uniform magnetic field in the domain introduces a
numerical issue: the simulation time step is determined by the low
density external medium and is $<1/100$ of that in the disk
midplane (region of primary interest). To speed up the simulation, we
artificially reduce the magnetosonic sound speed by adopting the Boris
method with a maximum magnetosonic sound speed of
$2500\:{\rm{km\:s}}^{-1}$ \citep[][]{Boris1970,
  Gombosi_ea2002}. Although this procedure is, in principle, only
valid for simulations of steady state problems, it is still applicable
when the flow velocities are much smaller than the sound speed.

Following methods of Paper I, to model star formation we allow
collisionless star cluster particles, i.e., point masses representing
star clusters or sub-clusters of minimum mass $M_*=100\:M_{\odot}$, to
form.  These particles are created when the density within a cell
exceeds a star formation threshold value of
$n_{\rm{H,sf}}=10^5\:{\rm{cm}}^{-3}$.  We assume a local SFR that
converts a fraction, $\epsilon_{\rm{ff}}=0.02$, of gas above the
threshold density into stars per local free-fall time
\citep[][]{KrumholzTan2007}.  No mass-to-flux ratio criterion is
included, so magnetically-subcritical cells are also able to form
stars, with necessary flux redistribution assumed to be occurring at
sub-grid scales.  Particle motions are calculated by using the
gravitational field interpolated from the grid to the particle
positions in the equation of motion. The mass of the particles is
included when solving the Poisson equation.
 
\section{Results}\label{sect:results}

\begin{figure*}
\begin{center}
\includegraphics[width=\textwidth]{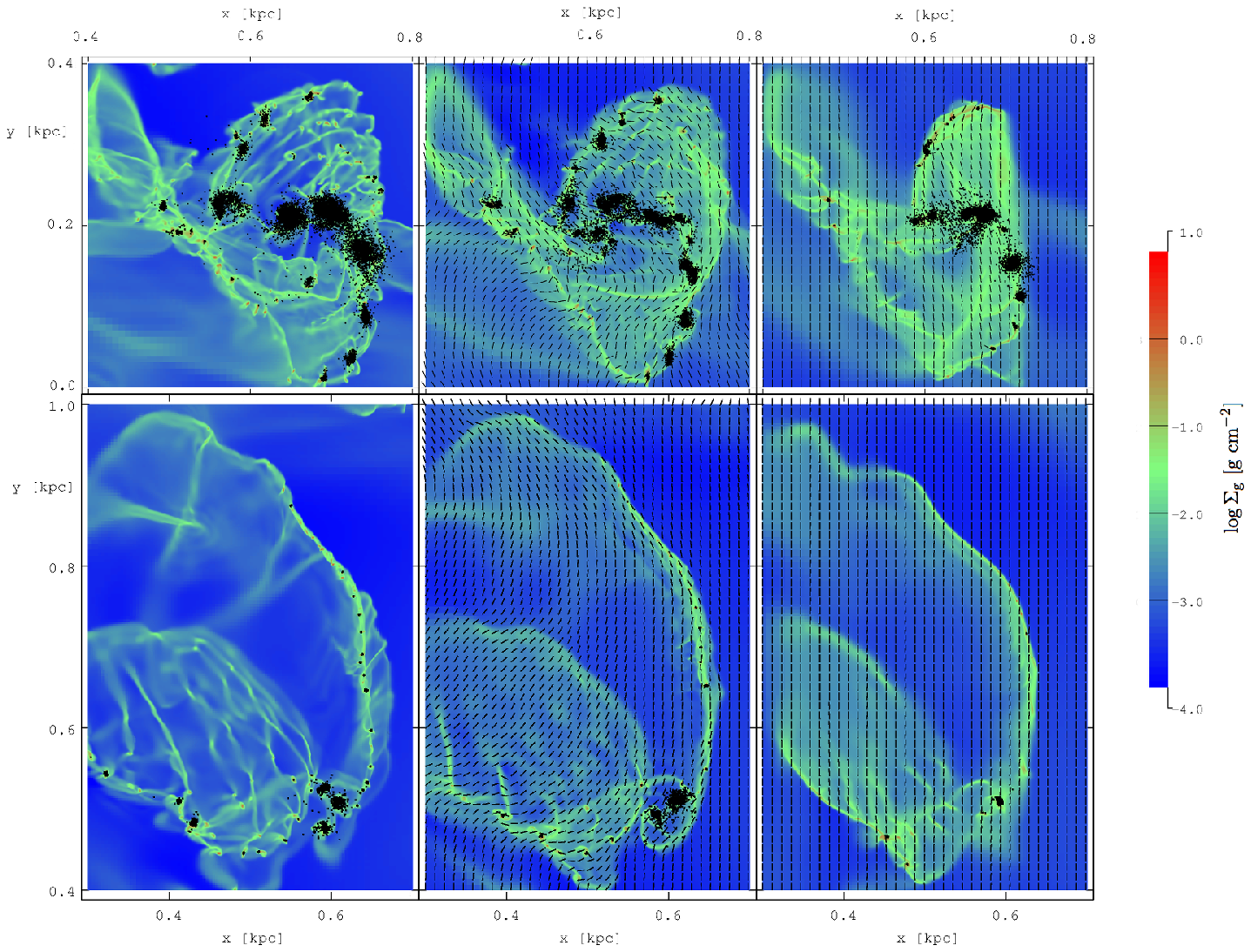}
\caption{
Mass surface densities of Region 1 (top) and 2 (bottom) for
$0\:{\rm{\mu}}$G (left), $10\:{\rm{\mu}}$G (middle) and
$80\:{\rm{\mu}}$G (right). Black dots show star cluster particles,
each representing $100\:M_\odot$ of stars, while lines indicate
direction of mass-weighted magnetic field.  }
\label{fig:regions}
\end{center}
\end{figure*}

\subsection{Cloud Structure and Fragmentation}\label{sect:cloud_structure}

Figure~\ref{fig:surface_density} shows mass surface densities and
temperatures within the numerical domain for models with different
initial magnetic field strengths. The pure hydrodynamical model has a
very fragmented structure, which becomes smoother as magnetic field
strength increases. Significant differences are also seen in the
temperature distributions, which are the result of different gas
densities (and thus different heating/cooling rates and
equilibrium temperatures) and also different amounts of shock heating
from large scale flows and accretion heating from dense clump
formation.  Figure~\ref{fig:regions} shows zoom-ins of two selected
regions, to illustrate more clearly the effects of magnetic field
strength on resulting gas structures. Differences in star formation
activity are discussed below (\S\ref{S:starform}).

The fragmentation can be quantified by counting the number of separate
clouds that form after 10~Myr of evolution.  As in Paper I, we assume
a threshold density of $n_{\rm{H}}=100\:{\rm{cm}^{-3}}$ to define gas
that is part of molecular ``cloud'' structures. For simplicity, all
connected cells are counted as a single cloud. Using this routine, we
find 287, 134 and 28 clouds in the $0,\:10$ and $80\:{\rm{\mu}}$G
models, respectively. The 
mean[median] cloud masses are 4.0$\times10^4\:[27]\:M_{\odot}$ for
$0\:{\rm{\mu}}$G, 7.3$\times10^4\:[100]\:M_\odot$ for
$10\:{\rm{\mu}}$G and 3.1$\times10^5\:[5.0\times10^4]\:M_\odot$ for
$80\:{\rm{\mu}}$G. Most clouds forming in the 0 and 10$\:\rm{\mu}$G
models are thus of very low mass. As expected, increasing the magnetic
field strength has a major impact on the fragmentation of the initial
GMCs. The magnetic critical mass of spherical clouds,
$M_B=16.2(B/10\:{\rm{\mu}G})^3(n_{\rm{H}}/1000\:{\rm{cm}^{-3}})^{-2}$
\citep[][]{BertoldiMcKee1992}, shows we expect a factor $\sim500$
change in typical cloud mass comparing 10 and 80$\:\rm{\mu}$G cases,
which is seen in the median cloud masses.

In Figures~\ref{fig:surface_density} and \ref{fig:regions} we also see
that the magnetic field is more distorted by gas motions in Region 1
than Region 2.
In the $10\:{\rm{\mu}}$G run, filaments predominantly lie parallel or
perpendicular to field orientation. On $\sim100\:{\rm{pc}}$ scales,
the 80$\:\rm{\mu}$G fields are harldly affected by the cloud motions,
but do become more distorted and disordered in regions of high star
formation activity.

\subsection{Column and Volume Density Distributions}

\begin{figure*}
\begin{center}
\includegraphics[width=\textwidth]{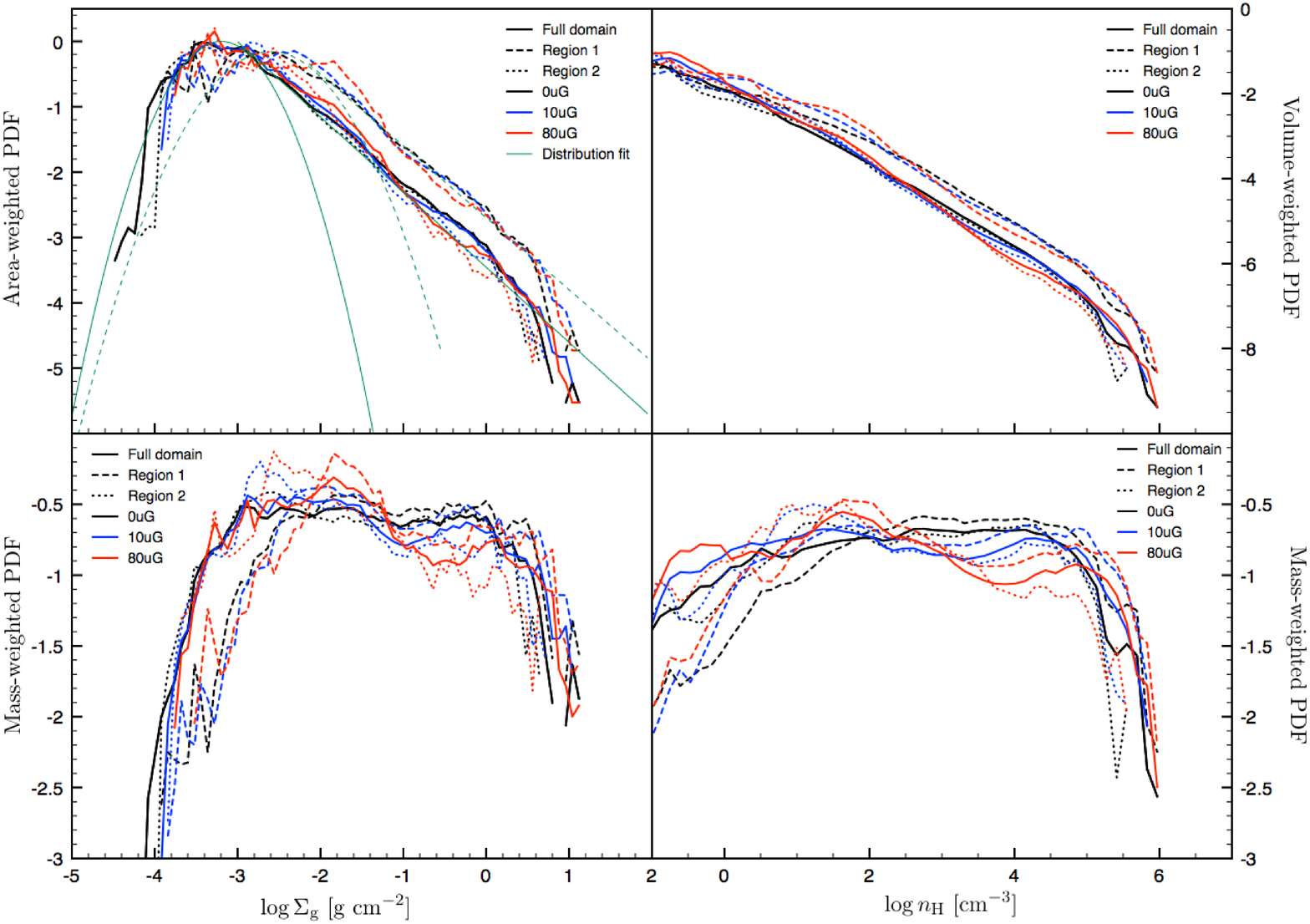}
\caption{
Mass surface density PDFs weighted by area (top left) and by mass
(bottom left) and volume density PDFs weighted by volume (top right)
and mass (bottom right). In each panel, 0, 10, 80$\:\rm{\mu}$G models
are shown with black, blue and red lines, respectively. Solid lines
show the whole kpc-sized region, dashed lines Region 1, and dotted
lines Region 2. Thin green lines in top left panel show log-normal and
power-law distributions fitted to these two PDFs.
}
\label{fig:PDFs}
\end{center}
\end{figure*}

We also quantify ISM structure by examining the probability
distribution functions (PDFs) of $\Sigma_g$ (as viewed from above the
disk) and number density of H nuclei,
$n_{\rm{H}}$. Figure~\ref{fig:PDFs} shows these quantities weighted by
area/volume (top row) and by mass (bottom row). The $\Sigma_g$ PDF is a
directly observable feature of clouds and galaxies, while the volume
density PDF needs to be reconstructed using different techniques
\citep[e.g.,][]{Kainulainen_ea2014}.

The $\Sigma_g$ PDFs show the typical combination of a log-normal with
power-law tail at high surface densities
\citep[e.g.,][]{Kainulainen_ea2009}.  The transition occurs around
$10^{-2}\:{\rm{g\:cm}^{-2}}$, corresponding roughly to the minimum
$\Sigma_g$ of regions containing ``cloud'' (GMC) gas, which is
essentially all part of self-gravitating structures. There is
variation from region to region, with Region 1 having its log-normal
peaking at higher $\Sigma_g$. The shape of the power-law distribution
differs only slightly between these individual regions, roughly
scaling as $\Sigma_{g}^{-\alpha_{\rm{\Sigma,PDF}}}$ with
$\alpha_{\rm{\Sigma,PDF}}\simeq1$ for the area-weighted PDF, and thus
relatively flat slopes of the mass-weighted distributions.  Power law
tail normalization at given $\Sigma_g$ is several times higher in
Region 1 than Region 2, i.e., the former has a higher dense gas mass
fraction. This also correlates with its more fragmented morphology and
larger SFR (below).

There are only modest differences seen in the PDFs as a function of
magnetic field strength (although note the large dynamic range in the
figures). At higher $\Sigma_g$, these are more clearly seen in Region
2, which also shows the greatest effect from $B$-field strength on its
degree of fragmentation and SFR (below).

Comparing to observations, somewhat steeper indices, i.e.,
$\alpha_{\rm{\Sigma,PDF}}\approx1.33-3.65$, have been reported in dust-emission-derived
$\Sigma_g$ PDFs by \citet[][]{Schneider_ea2014}. We note that some
extinction-based studies find PDFs that can be fit by a single
log-normal up to $\Sigma\simeq0.5\:{\rm{g\:cm}^{-2}}$
\citep[][]{Butler_ea2014}. Note, these observations are derived from
``in-plane'' higher-resolution views of clouds, in contrast to the
``top-down'' views of the galactic plane from the
simulations. Direct observational constraints of top-down views are
expected to be able to achieved with high angular resolution
observations of nearby galactic disks, e.g., with {\it{ALMA}}.

Volume density PDFs in the simulations show similar trends as the
$\Sigma_g$ PDFs, with Region 1 having a larger fraction of gas at
higher densities. For $n_{\rm{H}}\geq100\:{\rm{cm}^{-3}}$, i.e., the
cloud gas, the distributions can be reasonably well approximated with
single power laws up to $\sim10^5\:{\rm{cm}^{-3}}$ with power law
index of $\simeq-1$.

\subsection{Evolution of Dense Gas Mass Fractions \& SFRs}\label{S:timeevol}

\begin{figure*}
\begin{center}
\includegraphics[width=\textwidth]{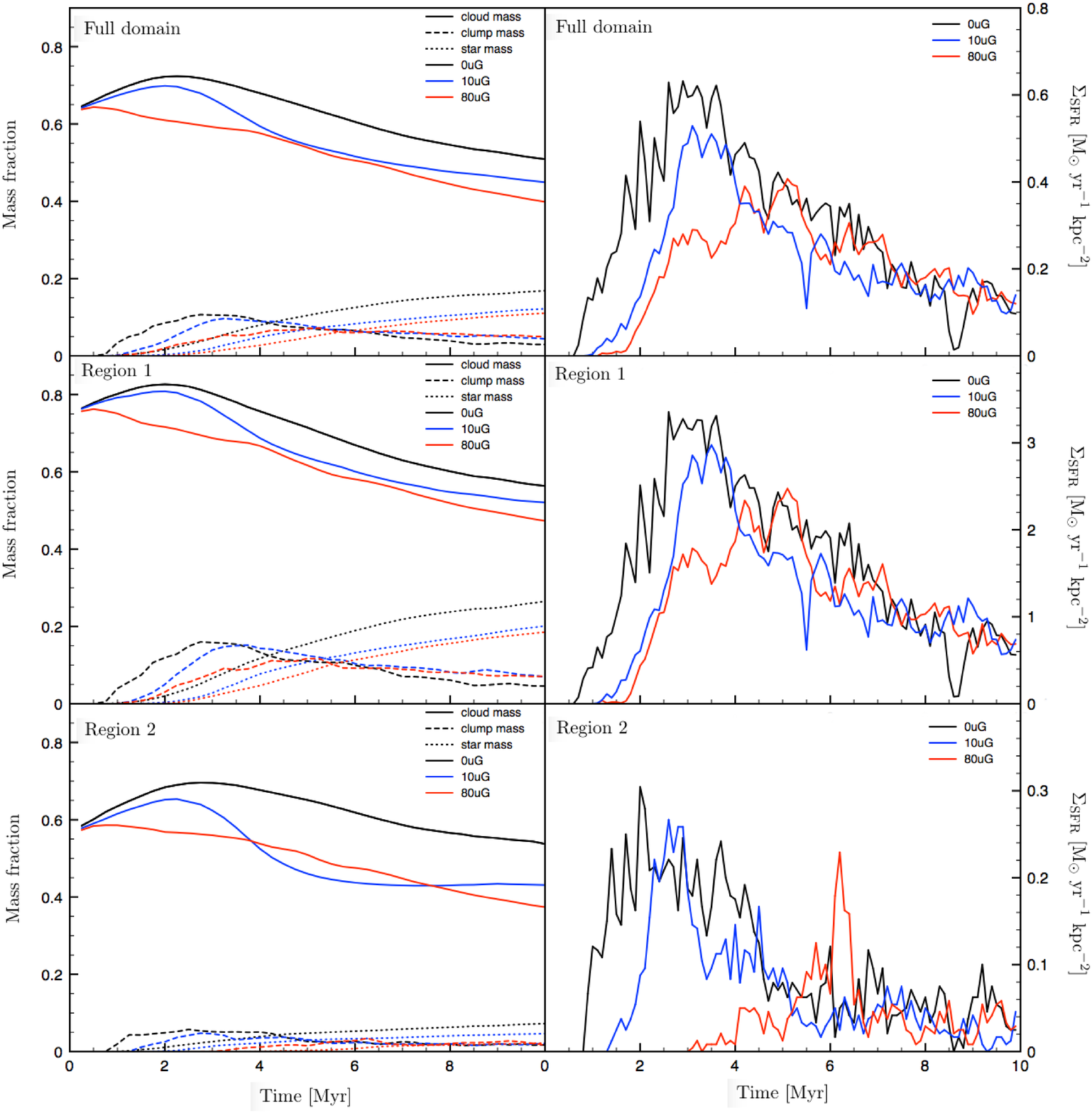}
\caption{
Time evolution of mass fraction (left) in clouds (solid), dense clumps
(dashed) and star cluster particles (dotted) and SFR per unit disk
area (right) for different values of magnetic field, i.e.,
$0\:{\rm{\mu}}$G (black), $10\:{\rm{\mu}}$G (blue) and
$80\:{\rm{\mu}}$G (red). Top panel is for full domain, middle for
Region~1 and bottom for Region~2.  }
\label{fig:massfraction}
\end{center}
\end{figure*}

Here we consider the time evolution of cloud
($n_{\rm{H}}>100\:{\rm{cm}}^{-3}$) and clump
($n_{\rm{H}}>10^5\:{\rm{cm}}^{-3}$) mass fractions over the
$10\:{\rm{Myr}}$ timescale of the simulations
(Fig.~\ref{fig:massfraction}). For the whole kpc-sized region, cloud
mass fractions start from relatively high values, $\simeq0.6$,
inherited from the TT09 simulation. Over $10\:{\rm{Myr}}$, they show a
moderate decline, partly because of the build up a mass fraction in
stars, and partly in the magnetized cases from expansion of cloud
envelopes due to the introduced magnetic pressure. Clump mass
fractions grow to peak values of $\sim0.05-0.1$, with smaller peak
values reached more slowly as $B$-field is increased. Since stars are
only allowed to form from clump gas, SFR evolution closely tracks
clump mass fraction evolution, although it exhibits more
stochasticity.  At late times the different models appear to converge
in their clump mass fractions and SFRs, perhaps due to the exhaustion
of most of the initially unstable gas via star formation.

Regions 1 and 2 follow similar trends. Indeed Region~1's area of
0.16$\:\rm{kpc}^2$ contains large fractions of the clump mass and star
formation of the whole kpc region. Region~2 starts with gas structures
that are somewhat easier to stabilize with magnetic fields (at least
in directions perpendicular to the initial $y$-direction field
orientation), so larger differences are seen in clump and SFR
evolution as field strength is increased. Region 1 contains $3\times$
as much cloud gas as Region~2. It contains two GMCs in the process of
merging, and the resulting cloud has significant kinetic energy. 
The magnetic field therefore plays a minor role in this case.
The overall SFR surface densities are about $10\times$ larger
in Region 1 compared to Region 2.

Some caveats are in order. First, the formation of clumps is a
response to simulation initial conditions, where dense,
self-gravitating clouds are allowed to collapse to high densities as
the resolution is suddenly increased.
Second, once formed, star particles contribute gravitationally, but
stellar feedback, implementation of which is much more uncertain and
numerically challenging, is not yet included in the simulation.

\subsection{Magnetic Fields and Average SFRs}\label{S:starform}

\begin{figure}
\begin{center}
\includegraphics[width=8cm]{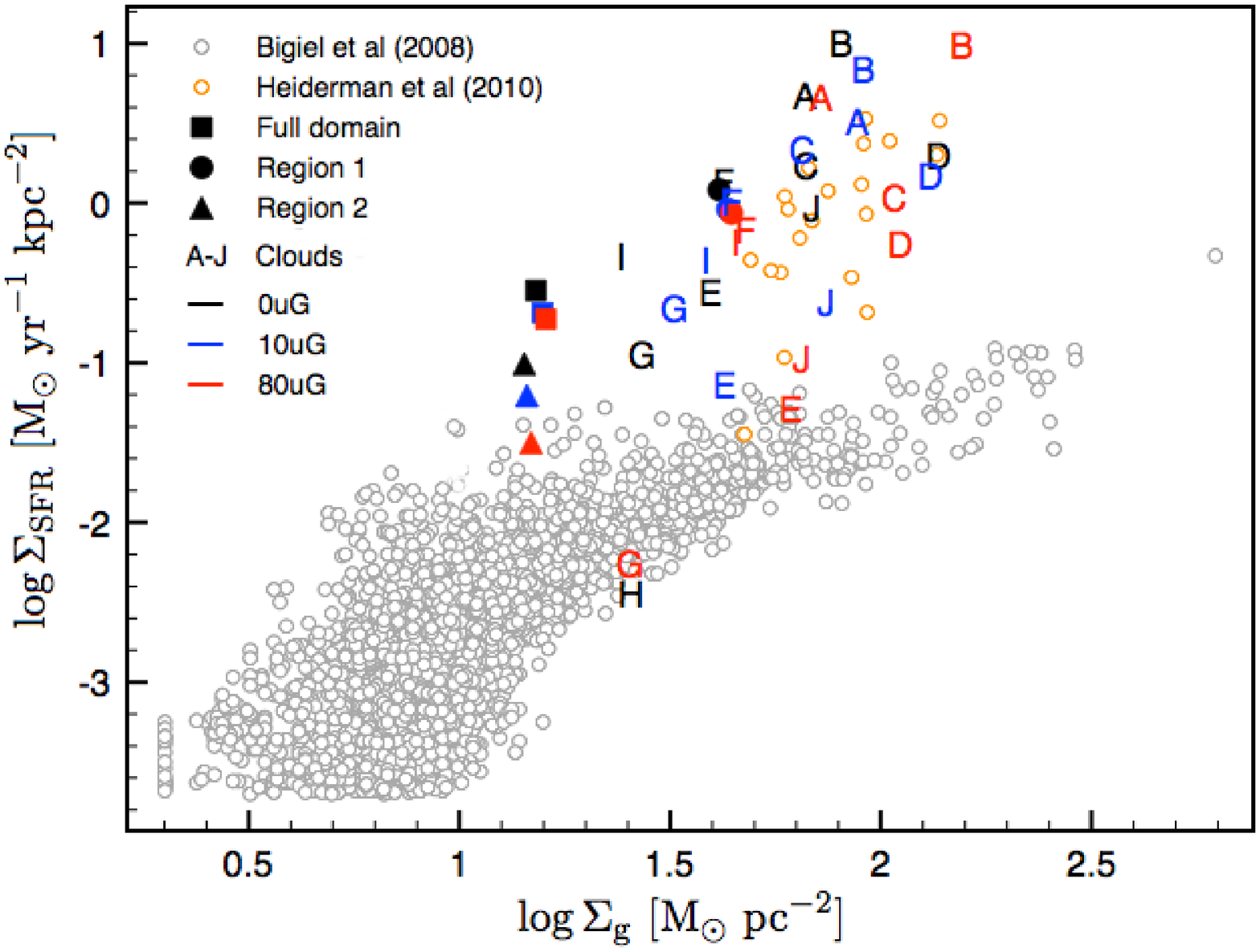}
\caption{
SFR surface density versus total gas mass surface density showing
kiloparsec-scale data \citep[][]{Bigiel_ea2008} (open grey circles)
and GMC data \citep[][]{Heiderman_ea2010} (open orange circles). Solid
symbols show mean values for full kpc domain (squares), Region 1
(circle) and Region 2 (triangle), while selected cloud regions are
indicated with the letter assigned in
Fig.~\ref{fig:surface_density}. Colors indicate magnetic field
strength: $0\:{\rm{\mu}}$G (black), $10\:{\rm{\mu}}$G (blue) and
$80\:{\rm{\mu}}$G (red).
}
\label{fig:KS}
\end{center}
\end{figure}

The detailed time history of clump and star formation is sensitive
to the choice of initial conditions. It is therefore useful to compare
the SFR averaged over the full $10\:{\rm{Myr}}$ evolution to the observations
since these are also averaged over $\sim 10\:{\rm{Myr}}$. Here we 
consider the absolute and relative average SFRs that are seen in the 
simulations on the
kiloparsec, ``region'' and ``cloud'' (i.e., $100\:{\rm{pc}}$) scales.

Figure~\ref{fig:KS} shows SFR surface density, $\Sigma_{\rm{SFR}}$,
versus $\Sigma_g$. Observational results are those of $\sim$kpc-scale
regions of galactic disks from \citet[][]{Bigiel_ea2008} and of
individual Galactic GMCs from \citet[][]{Heiderman_ea2010}.
On the kpc region scale, the simulations all start with
$\Sigma_g=17.0\:M_\odot\:{\rm{pc}}^{-2}$, decreasing to
$13.9,\:14.8,\:15.2\:M_\odot\:{\rm{pc}}^{-2}$ after 10~Myr for
$0,\:10,\:80\:\rm{\mu}$G cases, respectively, i.e.,
$\Sigma_{\rm{SFR}}=0.28,\:0.21,\:0.19\:M_\odot\:{\rm{yr^{-1}\:kpc^{-2}}}$. On
these scales, the magnetic field only has a modest impact on SFR, the
values of which are much higher ($\sim10-30\times$) than systems with
equivalent $\Sigma_g$ observed by
\citet[][]{Bigiel_ea2008}\footnote{The average SFR derived in Paper~I using {\it{Enzo}} is $2.8\times$
the equivalent hydrodynamic run with {\it{MG}}, which we attribute to
our implementation of improved methods for staggering introduction of
AMR at early times in the simulation.}.
However, as noted in \S\ref{S:timeevol}, total SFR is dominated by
that occurring in Region 1. When separate results for Regions 1 and 2
are considered, we see more significant effects of increasing the
magnetic field strength: a factor of 3 reduction in Region 2
(Fig.~\ref{fig:KS}).

Focusing on smaller $100\:{\rm{pc}}$-size ``Cloud'' scales (see
Fig.~1), we find simulated SFRs overlap with observational rates in
Galactic GMCs derived by \citet[][]{Heiderman_ea2010}.  Stronger
magnetic fields again almost always lead to lower SFRs.  One reason
for the decrease in SFR with increasing $B$-field is that clumps form
later in models with stronger fields (\S\ref{S:timeevol}).
In one case, Cloud H, stars only form in the $B=0\:{\rm{\mu}G}$ run.

Our simulations thus reproduce SFRs similar to some observed local
Galactic GMCs. Note the Heiderman et al. GMCs do not contain
especially vigorous regions of massive star formation, so may be
relatively less affected by internal star formation feedback.  On the
kpc scale, even our strongly magnetized simulation has values of
$\Sigma_{\rm{SFR}}$ that are too large. However, this is largely a
consequence of the ``starburst'' of Region 1. Inclusion of star
formation feedback is expected to have an impact in reducing this
activity, to be investigated in a future paper.

\section{Summary and Discussion}\label{sect:summary}

We have studied the effects of magnetic fields on molecular clouds
extracted from a global galaxy simulation, especially for cloud
structure and SFRs.  Magnetic fields suppress fragmentation, as
expected from consideration of the magnetic critical mass. Their
effects on $\Sigma_g$ and $n_{\rm{H}}$ PDFs are more modest.

In the context of models of star formation from gas above a threshold
density ($n_{\rm{H}}\geq10^5\:{\rm{cm}^{-3}}$), we find that on the
largest kpc scale, average SFRs are only modestly suppressed by
magnetic fields. However, this is due to the presence of a
starbursting region (Region 1) in the simulation domain. Considering
other regions, including down to ``GMC''-scales of $\sim100\:$pc, we
find a variety of $\Sigma_{\rm{SFR}}$s, with values quite similar to
those of some Galactic GMCs. Average suppression factors of
$\epsilon_B=0.76,\:0.54$ for $B=10,\:80\:{\rm{\mu}}$G compared to the
nonmagnetized case are seen in the clouds, but in some cases there can
be much more dramatic effects, including complete suppression in one
cloud.

As a numerical experiment, this study highlights several
important aspects, including issues of $B$-field initialization when 
overdense structures are already present in initial conditions. Stochastic 
effects are important, i.e., variation in behavior of individual GMCs, 
and kpc-scale SFRs can be influenced by a single GMC complex.
A sub-grid model for star formation is needed: here via a chosen
threshold density and an empirically motivated efficiency per local
free-fall time, $\epsilon_{\rm{ff}}=0.02$. Such low efficiencies may
require the effects of local star formation feedback, like
protostellar outflows, to maintain turbulence and prolong star cluster
formation \citep[][]{NakamuraLi2007}. And/or they may result from the
influence of magnetic fields themselves.
Note, we have not included magnetic field effects on the star formation 
sub-grid model, rather keeping its empirically-based parameters fixed 
for simplicity. There is scope in future work for exploring magnetic 
sub-grid star formation models, where cell mass-to-flux ratio is also 
an input, which should lead to a more natural threshold criterion for 
star formation activity.

Other effects of local star formation feedback, such as stellar winds,
ionization and supernovae, have not yet been included in these
simulations, but are expected to act to further reduce SFRs. To
disentangle the relative importance of these effects and of magnetic
fields will require testing many properties of the simulated clouds and
young stellar populations against observed systems
\citep[e.g.,][]{Butler_ea2014b}.

\acknowledgements SvL acknowledges support from the SMA Postdoctoral
Fellowship (SAO). JCT acknowledges support from NASA grant
ATP09-0094. Resources supporting this work were provided by NASA
High-End Computing Program, the {\it Smithsonian Institution High
  Performance Cluster} and the High Performance Computing facilities
at the University of Leeds.

\end{document}